\documentclass{JHEP3} 

\usepackage{epsfig,multicol}
\usepackage{amsmath,amssymb}
\usepackage{epic}
\usepackage{concmath,palatino}

\newcommand\fverb{\setbox\pippobox=\hbox\bgroup\verb}
\newcommand\fverbdo{\egroup\medskip\noindent%
			\fbox{\unhbox\pippobox}\ }
\newcommand\fverbit{\egroup\item[\fbox{\unhbox\pippobox}]}
\newbox\pippobox

\newcommand{\be}{\begin{equation}} 
\newcommand{\ee}{\end{equation}}

\newcommand{\ba}{\begin{eqnarray}}
\newcommand{\ea}{\end{eqnarray}}

\newcommand{\ads}{AdS_5\times S^5}

\newcommand{\refeq}[1]{Eq.~(\ref{#1})}

\title{Large spin expansion of the long-range Baxter equation in the $\mathfrak{sl}(2)$ sector of ${\cal N}=4$ SYM}

\author{Matteo Beccaria and Francesca Catino\\
  Dipartimento di Fisica, Universita' del Salento,
  Via Arnesano, 73100 Lecce\\
  INFN, Sezione di Lecce\\
  E-mail: \email{matteo.beccaria@le.infn.it}, \\ \phantom{E-mail:} \email{francesca.catino@le.infn.it}}

\abstract{
Recently, several multi-loop conjectures have been proposed for the spin dependence of 
anomalous dimensions of twist-2 and 3 operators in the $\mathfrak{sl}(2)$ sector of ${\cal N}=4$ SYM. 
Currently, these conjectures are not proven, although several consistency checks have been performed 
on their large spin expansion. In this paper, we show how these expansions can be efficiently computed 
without resorting to any conjecture. To this aim we present in full details a method to expand at large spin
the solution of the long-range 
Baxter equation. We treat the twist-2 and 3 cases at two loops and the twist-3 case at three loops.
Several subtleties arise whose resolution leads to a simple algorithm computing the expansion.
}

\begin{document} 

\section{Introduction}
\label{sec:Intro}

Integrability of four dimensional {\em planar} Yang-Mills theories is a quite relevant feature that permits in principle
to obtain non-perturbative results~\cite{Belitsky:2004cz}. It first appeared as a surprising fact in the one-loop analysis of 
special QCD subsectors~\cite{QCD-int}. Later, its universality became more and more clear due to the many 
extensions at higher orders~\cite{Belitsky:2004sc,Belitsky:2004sf,Belitsky:2005bu}.

Currently, a deep understanding of its origin is achieved by means of Maldacena AdS/CFT correspondence~\cite{Maldacena:1997re}.
In the maximal supersymmetric ${\cal N}=4$ Yang-Mills theory, integrability of the superconformal side is related to integrability
of the dual superstring theory on $\ads$~\cite{Bena:2003wd}. As a major outcome, a deeply motivated higher loop proposal for the $S$-matrix of 
the pair ${\cal N}=4$ SYM/type II on $\ads$ is now 
available~\cite{Arutyunov:2004vx,Staudacher:2004tk,Beisert:2005fw,Beisert:2005tm,Janik:2006dc,Eden:2006rx,Hernandez:2006tk,Plefka:2006ze,Beisert:2006ez,Beisert:2006ib}.
In this paper, we shall work within this framework assuming the validity of the above Ansatz, although the subject is still under
development~\cite{Arutyunov:2007tc}.

We restrict our considerations to the so-called $\mathfrak{sl}(2)$ sector of ${\cal N}=4$ SYM
which is an invariant  subsector closed under perturbative renormalization mixing. It is spanned 
by single trace operators obtained as linear combinations of the basic objects
\be 
\label{eq:sl2states}
\mbox{Tr}\left( {\cal D}^{n_1}\,\varphi \cdots {\cal D}^{n_L}\,\varphi\right),\quad n_1 + \cdots + n_L = N,
\ee 
where $\varphi$ is one of the three complex scalar fields of ${\cal N}=4$ SYM and ${\cal D}$ is a light-cone covariant
derivative. The numbers $\{n_i\}$ are non-negative integers and their sum $N$ is the total spin. As usual, 
the number $L$ of $\varphi$ fields is called the {\em twist} of the operator. It equals the classical dimension minus the spin.
The subsector of states with fixed spin and twist is also perturbatively closed under renormalization mixing.

The $\mathfrak{sl}(2)$ sector is very rich and interesting.
Even when the twist is kept low, it is a non-trivial infinite dimensional sector with certain similarities with 
analogous QCD composite operators appearing in deep inelastic scattering.

At one loop, the dilatation operator can be interpreted as the integrable Hamiltonian of a spin chain with $L$ sites. At each site, 
the degrees of freedom, associated with ${\cal D}^n\,\varphi$, transform in the $s=1/2$ infinite dimensional $\mathfrak{sl}(2)$ 
representation~\cite{Beisert:2003yb}.

Going beyond one loop, asymptotic all-order Bethe Ansatz equations have been proposed~\cite{Staudacher:2004tk}. Important
successful checks are described in~\cite{Eden:2005bt,Zwiebel:2005er,Eden:2006rx}.
Thanks to supersymmetry, wrapping problems (see~\cite{Kotikov:2007cy,Janik:2007wt} for recent developments) 
only occur at $L+2$ loop order in twist-$L$ operators~\cite{Beisert:2005fw,Beisert:2005tm}.
Hence, if we are interested in a three loop analysis, the twists $L=2$ and 3 are safe. We shall focus on these cases.

Scaling composite operators are elements in the span of \refeq{eq:sl2states} which are 
eigenvectors of the dilatation operator. The eigenvalues
are the anomalous dimensions $\gamma(N, L, \lambda)$ where $\lambda$ is the 't Hooft planar coupling 
($N_c\to\infty$ is the number of colors sent to infinity with constant $\lambda$) 
\be
\lambda = \frac{g_{\rm YM}^2\,N_c}{8\,\pi^2},
\ee 
At fixed twist $L$, 
one is interested in the perturbative expansion of $\gamma(N, L, \lambda)$ that takes the form 
\be
\gamma = \sum_{n=1}^\infty \gamma_n(N)\,\lambda^n.
\ee
Technically, integrability permits to write down Bethe Ansatz equations that compute $\gamma$ order by order in $\lambda$. The Bethe equations
are algebraic equations for the Bethe roots. Given a solution, one inserts the Bethe roots in a simple closed expression providing  $\gamma$. 
The number of Bethe roots is precisely $N$. So, the above procedure must be repeated at each given $N$. It is not 
obvious how to find a parametric expression of $\gamma_n(N)$ as a function of $N$.
Nevertheless, such closed formulae are important in physical applications like, for instance, 
the BFKL analysis of pomeron exchange~\cite{Lipatov:1976zz}.

Recently, work on twist-2 and 3 operators has led to higher order conjectures
for the functions $\gamma_n(N)$ in various twists and in generalized subsectors~\cite{Kotikov:2007cy,Beccaria:2007pb,Beccaria:2007bb,Beccaria:2007vh,Beccaria:2007cn}.
However, proofs are missing, at least beyond one-loop.

The general way to deduce such conjectures is somewhat deceiving. One starts by solving the Bethe Ansatz equations at various $N$. Generally, the Bethe roots are
non-trivial algebraic numbers. Nevertheless, $\gamma_n(N)$ turns out to be a rational number to any precision. Given a sufficiently long sequence of such rational 
numbers, one makes an inspired guess about the closed formula for $\gamma_n(N)$. General arguments from Feynman diagram calculations and deep properties like
reciprocity~\cite{Basso:2006nk,Dokshitzer:2006nm,Beccaria:2007bb} are invoked to constrain the general expression. In several cases, a simple expression is found.

However, if it were not for the amount of {\em inspiration}, these conjectures would be nothing but interpolation formulae, although surprisingly simple. 
It would be very nice to prove these formulae without making conjectures. Three {\em natural} lines of investigation are the following:
\begin{enumerate}
\item {\em Bethe Ansatz equations}. As we mentioned, they deal with the Bethe roots as the basic object. This is 
a bit unnatural if one is ultimately interested in the anomalous dimensions which are a much simpler quantity. 
The Bethe equations can be used to obtain the asymptotic density of Bethe roots, {\em i.e.} the $N\to\infty$ 
limit~\cite{Eden:2006rx,Beisert:2006ez,Rej:2007vm}. However, 
systematic corrections at large $N$ appear to be rather involved and practically restricted to a few orders~\cite{Casteill:2007ct}.

\item {\em Baxter equation}. An alternative approach to the calculation of $\gamma(N)$ is based on the Baxter approach originally formulated in~\cite{Bax72}.
The crucial ingredient is the Baxter operator whose eigenvalues $Q(u)$ obey a relatively simple functional equation.
If $Q(u)$ is assumed to be a polynomial, then the Baxter equation is equivalent to the algebraic Bethe Ansatz equations for its roots
to be identified with the Bethe roots. A more general discussion with references to the non-polynomial cases can be found 
in~\cite{Derkachov:1999pz,KorTrick1}.
Remarkably, the polynomial $Q(u)$ turns out to have rational coefficients in all cases where a closed Ansatz for $\gamma(N)$ has been proposed.
Accordingly, the anomalous dimension is a rational combination of derivatives of $Q(u)$. This approach leads to the known
exact one-loop formulae. The key feature is that the one-loop Baxter polynomial is known as an analytic  function of $N$ in this case.
Unfortunately, such a nice result is not available  beyond one-loop.

\item {\em Conformal methods}. These are reviewed in the recent paper~\cite{Belitsky:2007jp}.
Leading twist conformal primary operators lie on the unitarity bound and hence are conserved (irreducible) in the free theory. 
It is possible to cleverly exploit the pattern of breaking of the irreducibility conditions in the interacting theory. In the end, it is possible 
to gain an order of perturbation theory and infer two-loop results from the lowest order calculations.
\end{enumerate}

In this paper, we shall reconsider the Baxter approach. We give up the task of computing the exact Baxter polynomial 
at more than one loop parametrically in the spin $N$. Instead, we develop a technique to systematically derive the large spin expansion of $\gamma(N)$
at high orders in the natural logarithmic expansion $\sim \log^n\,N/N^m$.
This will permit to derive from first principles the expansion of $\gamma(N)$ although without knowing its resummation. This can be 
regarded as a strong check of the proposed conjectures at large spin. It is also a useful result by itself if, for instance, one is interested in 
proving large spin properties like reciprocity.

The sketch of the paper is the following. First, we illustrate the method in the easy one-loop case in twist-2 and 3. At this level, we do nothing more than 
rephrase a very tricky technique originally devised by G.~Korchemsky in~\cite{KorTrick1}. The resulting procedure will be called $\Delta$-method for brevity.
It works well and the desired expansion is systematically obtained 
in a few lines of calculations, easily implemented on symbolic algebra packages.

Then, we move to the two-loop case. Here, we find some surprise. The $\Delta$-method fails to reproduce the rational part of $\gamma_2(N)$ starting from the terms 
${\cal O}(1/N^4)$. Curiously, the failure is not signaled by any apparent pathology. The lesson is that the $\Delta$-method can be 
misleading, although it works well for the non rational part and for all the logarithmically enhanced contributions. The problem is not present in twist-3
as we illustrate and partially explain.

As a further step, we analyze what is happening in twist-2 and propose a safe improved expansion of the Baxter equation which correctly reproduces the full
anomalous dimension. Finally, we analyze the twist-3 case at three loop to show that again there is a failure in the rational part of $\gamma$.
The same solution adopted for the twist-2, 2-loop case, solves completely the problem and appear to be the necessary universal recipe for this kind of 
expansions.

\section{The one-loop Baxter equation in the $\mathfrak{sl}(2)$ sector}

\subsection{General structure of the Baxter equation}
 
In the notation of~\cite{Belitsky:2006en}, the one-loop Baxter equation in $\mathfrak{sl}(2)$-like sectors is 
\be
(u+i\,s)^L\,Q(u+i) + (u-i\,s)^L\,Q(u-i) = t_L(u)\,Q(u),
\ee
where $L$ is the twist, $s$ the conformal spin, and $t_L(u)$ a polynomial 
\be
t_L(u) = 2\,u^L + q_{L, 2}\,u^{L-2} + \cdots + q_{L, L}.
\ee
For brevity, in the following, we shall refer to $t_L(u)$ as the {\em transfer matrix} with a small abuse of language.

\medskip
The second charge $q_{L, 2}$ is explicitly known and reads
\be
q_{L, 2} = -(N+L\,s)(N+L\,s-1)+L\,s\,(s-1),
\ee
where $N$ is the spin quantum number of the solution. We shall be interested in the polynomial solution
\be
Q(u) = \prod_{i=1}^N (u-u_i).
\ee
Replacing into the Baxter equation, we obtain the Bethe Ansatz equations for the $XXX_s$ $\mathfrak{sl}(2)$ spin chain
\be
\left(\frac{u_k+i\,s}{u_k-i\,s}\right)^L = \prod_{j=1, j\neq k}^N\frac{u_k-u_j-i}{u_k-u_j+i}.
\ee
The one-loop anomalous dimension/energy and quasi-momentum are given in terms of the Bethe roots $u_i$ by 
\be
\gamma_1 = \sum_{k=1}^N\frac{2\,s}{u_k^2+s^2},\qquad
e^{i\,\theta} = \sum_{k=1}^N\frac{u_k-i\,s}{u_k+i\,s}.
\ee
They can be calculated from the Baxter polynomial, completely bypassing the knowledge of $\{u_i\}$
\be
\gamma_1 = i\,\left. (\log Q(u))' \vphantom{\frac{1}{2}} \right|_{u=-i\,s}^{u=+i\,s},\qquad 
e^{i\,\theta} = \frac{Q(+i\,s)}{Q(-i\,s)}.
\ee
In the $\mathfrak{sl}(2)$ sector, we must set $s=\frac{1}{2}$.
Also we shall always take $N$ {\em even}. In this case, the ground state with minimal anomalous dimension is non degenerate
and has $Q(u) = Q(-u)$ with automatically vanishing quasi-momentum. This is correct for single-trace composite
operators in the gauge theory. Exploiting parity, the anomalous dimension is simply
\be
\gamma_1 = 2\,i\,\left. (\log Q(u))' \vphantom{\frac{1}{2}} \right|_{u=+i\,s}.
\ee

\subsection{Ground state solution in twist 2 and 3}

The structure of the ground state, {\em i.e.} with minimal anomalous dimension, is rather simple for even $N$ and twist 2 or 3.
In the twist-2 case, $t_2(u)$ does not depend on any unknown quantum number
\ba
t_2(u) &=& 2\,u^2 + q_{2, 2},\\
q_{2,2} &=& -\frac{1}{2}-N\,(N+1).
\ea
The Baxter function is the even hypergeometric polynomial~\cite{Derkachov:2002tf}
\be
Q(u) = {}_3\,F_2\left(-N, N+1, \frac{1}{2}-i\,u; 1, 1; 1\right).
\ee
From this explicit expression we can compute the one-loop anomalous dimension
\be
\gamma_1(N) = 4\,S_1(N),
\ee
where (nested) harmonic sums are defined as usual by
\be
S_a(N) = \sum_{n=1}^N\frac{(\mbox{sign}\,a)^n}{n^{|a|}},\qquad
S_{a, \mathbf{b}} = \sum_{n=1}^N\frac{(\mbox{sign}\,a)^n}{n^{|a|}}\,S_{\mathbf{b}}(n).
\ee
In particular $S_1$ is the harmonic sum
\be
S_1(N) = 1 + \frac{1}{2} + \cdots + \frac{1}{N}.
\ee

The twist-3 case is also simple because again $t_3(u)$ do not depend on any unknown quantum number. 
\ba
t_3(u) &=& 2\,u^3 + q_{3, 2}\,u,\\
q_{3,2} &=& -\frac{3}{4}-\left(N+\frac{1}{2}\right)\,\left(N+\frac{3}{2}\right).
\ea
The Baxter function  is the even hypergeometric polynomial~\cite{Beccaria:2007cn,Beccaria:2007bb,Kotikov:2007cy} 
\be
Q(u) = {}_4\,F_3\left(-\frac{N}{2}, \frac{N}{2}+1, \frac{1}{2}+i\,u, \frac{1}{2}-i\,u; 1, 1, 1; 1\right),
\ee
which can be written in terms of Wilson polynomials (see Appendix B of~\cite{KorTrick2}).

From this explicit expression we can compute the one-loop anomalous dimension
\be
\gamma_1(N) = 4\,S_1\left(\frac{N}{2}\right).
\ee

\subsection{Large $N$ expansion of the exact $\gamma_1$}

The large $N$ expansion of $\gamma_1$ is that of the function $4\,S_1(1/\varepsilon)$ for small values of the variable
$\varepsilon$ defined as
\be
\mbox{twist-2}\quad \varepsilon=\frac{1}{N},\qquad
\mbox{twist-3}\quad \varepsilon=\frac{2}{N}.
\ee
The expansion is well-known in analytic form at all orders and reads
\be
\label{eq:oneloopexp}
4\,S_1\left(\frac{1}{\varepsilon}\right) = -4\,\log\varepsilon+4\,\gamma_E+2\,\varepsilon-2\,\sum_{k=1}^\infty\frac{B_{2\,k}}{k}\,\varepsilon^{2\,k},
\ee
where $B_k$ are Bernoulli numbers defined by 
\be
\frac{t}{e^t-1} = \sum_{k=0}^\infty \frac{B_k}{k!}\,t^k = 1-\frac{t}{2}+\frac{t^2}{12}-\frac{t^4}{720}+\frac{t^6}{30240}-\frac{t^8}{1209600}+{\cal O}\left(t^{10}\right).
\ee

\subsection{Large $N$ expansion from the Baxter equation}

Now, our aim is to obtain the expansion \refeq{eq:oneloopexp} directly from the Baxter equation and without exploiting the known exact 
solution. Let us show that this is quite simple. We use a trick first introduced by G. Korchemsky~\cite{KorTrick1,KorTrick2} with some small 
modification to easy the implementation on symbolic calculation packages.

First, we define $u=i\,z$ and 
\be
Q(i\,z) = e^{F(z)},\qquad \Delta(z) = F(z+1)-F(z).
\ee
The cases twist-2 and 3 must be treated separately because the calculations differ in the details.

\subsubsection{Twist-2}

We write the Baxter equation as
\be
\label{eq:bxt:one:twist2}
\varepsilon^2\,\left(z+\frac{1}{2}\right)^2\,e^{\Delta(z)}+\varepsilon^2\,\left(z-\frac{1}{2}\right)^2\,e^{-\Delta(z-1)} = 1+\varepsilon+\varepsilon^2\,\left(
\frac{1}{2}+2\,z^2\right).
\ee
The analysis of~\cite{KorTrick1} can be rephrased as the following very simple Ansatz for the asymptotic expansion of $\Delta(z)$
\be
\label{eq:delta}
\Delta(z) = -2\,\log\varepsilon + \sum_{n=0}^\infty a_n(z)\,\varepsilon^n.
\ee
This expansion is derived for sufficiently large positive $z$ and is assumed to admit an analytic continuation down to $z=1/2$.

\medskip
\noindent
Now, the procedure that we shall call $\Delta$-method goes in 3 simple steps:
\begin{enumerate}
\item Replace \refeq{eq:delta} into \refeq{eq:bxt:one:twist2} and match the various integer powers of $\varepsilon$ to
obtain the functions $a_n(z)$.

\item Given $\Delta(z)$, get back the derivative of $\log Q$, {\em i.e.} $F(z)$.

\item Evaluate $F'(1/2)$ to get the one-loop anomalous dimension.
\end{enumerate}
Let us see how all this can be accomplished. First, step (1). After expansion, we get
\ba
\Delta(z) &=& -2\,\log\,\varepsilon-2\,\log\left(z+\frac{1}{2}\right)+\varepsilon+2\,z^2\,\varepsilon^2 
-\left(2\,z^2+\frac{1}{6}\right)\,\varepsilon^3 +  \\
&& -\frac{1}{16}\, \left(4\, z^2+1\right) \left(12\, z^2-8\, z-1\right) \,\varepsilon^4
+\frac{1}{40}\, \left(240\, z^4-160\, z^3+120\, z^2-40\, z+3\right) \,\varepsilon^5 + \nonumber\\
&& +\frac{1}{12}\, \left(80 \,z^6-144\, z^5+120\, z^4-96 \,z^3+21\, z^2-3\, z\right) \,\varepsilon^6 + \nonumber\\
&& +\frac{1}{112}\, \left(-2240\, z^6+4032\, z^5-5040\, z^4+3808\, z^3-1540\, z^2+364\, z-33\right) \,\varepsilon^7+\cdots . \nonumber
\ea
To perform step (2), we first notice that from the definition of $\Delta(z)$ we deduce
\be
F'(z) = \frac{D}{e^D-1}\,\Delta(z).
\ee
The linear operator on the r.h.s. can be evaluated in closed form when acting on many special $g(z)$. In particular, this is true for all $g(z)$ that 
will appear in this and later calculations. The simplest case is that of polynomial $g(z)$
\be
\frac{D}{e^D-1} z^n = B_n(z),
\ee
where $B_n(z)$ is the $n$-th Bernoulli polynomial defined by the generating function
\be
\frac{t\,e^{t\,z}}{e^t-1} = \sum_{k=0}^\infty \frac{B_k(z)}{k!}\,t^k.
\ee
Also, for $g(z) = \log\,z$, we have the very useful identity
\be
\frac{D}{e^D-1} \,\log\,z = \psi(z),\qquad \mbox{where}\qquad\psi(z) = \frac{d}{dz}\,\log\Gamma(z).
\ee
Taking derivatives, we also obtain
\be
\frac{D}{e^D-1} \,\frac{1}{z^n} = \frac{(-1)^{n+1}}{(n-1)!}\,\psi^{(n)}(z),\qquad \psi^{(n)}(z) = \frac{d^n}{dz^n}\,\psi(z).
\ee
Going back to the problem of determining $F'(z)$, we immediately find
\ba
F'(z) &=& 
-2 \,\log \,\varepsilon -2\, \psi\left(z+\frac{1}{2}\right)+\varepsilon +\frac{1}{3} \,\left(6 \,z^2-6\, z+1\right) \,\varepsilon^2-\frac{1}{2}\, (2\, z-1)^2 \,\varepsilon^3+
\nonumber\\
&&
\frac{1}{240} \,\left(-720\, z^4+1920\, z^3-1560 \,z^2+480\, z-41\right)\, \varepsilon^4+\frac{1}{8} (2 \,z-1)^3 (6 \,z-7) \varepsilon^5+\nonumber\\
&&
+\frac{1}{252}\, \left(1680 \,z^6-8064   \,z^5+14280 \,z^4-12096 \,z^3+5145 \,z^2-1008 \,z+61\right)\, \varepsilon^6+\nonumber\\
&&
-\frac{1}{16} \,\left((2\, z-3)\, (2 \,z-1)^3 \,\left(20\, z^2-36\, z+17\right)\right)\, \varepsilon^7+\cdots.
\ea
The last step (3) is trivial. Using $\psi(1)=-\gamma_E$ to evaluate $\gamma_1 = 2\,F'(1/2)$ we 
immediately recover the correct expansion in full agreement with \refeq{eq:oneloopexp} !

\subsubsection{Twist-3}

The same analysis can be repeated in twist-3. The details are quite similar, but slightly different. The final expression of the function $\Delta(z)$ is now
\ba
\Delta(z) &=& -2\,\log\,\varepsilon + \log\,z-3 \log \left(z+\frac{1}{2}\right)+2 \log\,2+\,\varepsilon +\left(\frac{z^2}{2}-\frac{1}{8}\right) \,\varepsilon
   ^2+\left(-\frac{z^2}{2}-\frac{1}{24}\right) \,\varepsilon ^3+\nonumber\\
&& \left(-\frac{3 z^4}{16}+\frac{z^3}{8}+\frac{13 z^2}{64}+\frac{3 z}{64}-\frac{1}{1024
   (z-1)}+\frac{11}{256}+\frac{1}{1024 z}\right) \,\varepsilon ^4+\nonumber\\
&& \left(\frac{3 z^4}{8}-\frac{z^3}{4}+\frac{3 z^2}{32}-\frac{3 z}{32}+\frac{1}{512
   (z-1)}-\frac{7}{640}-\frac{1}{512 z}\right) \,\varepsilon ^5+\\
&& \left(\frac{5 z^6}{48}-\frac{3 z^5}{16}-\frac{5 z^4}{32}+\frac{z^3}{16}-\frac{z^2}{8}+\frac{9
   z}{128}-\frac{7}{4096 (z-1)}-\frac{7}{1536}+\frac{7}{4096 z}\right) \,\varepsilon ^6+\cdots . \nonumber
\ea
Again, each term can be worked out explicitly to give a simple analytic expression for $F'(z)$. Although $\Delta(z)$ is quite different than that for twist-2, 
one checks that again the same and correct expansion of $\gamma_1$ (now with $\varepsilon = 2/N$) is obtained.

\subsection{Comments}

Before moving to  the two-loop case, it is convenient to stop and comment about the above procedure. 

The $\Delta$-method in its original form is based on the following observation. The l.h.s. of the Baxter equation must balance the transfer matrix $t(u)$
at large $N$. This is accomplished for $\mbox{Im}\,u>0$ by assuming that the term $\sim (u+i/2)^L$ is dominant and treating the second piece as a perturbation.
This seems to be quite reasonable since in the end we want to compute $F'(z)$ in $z=1/2$ where the second piece vanishes. Our implementation starting directly from 
a logarithmic expansion of $\Delta(z)$ follows this simple idea. Notice however that the expansion is guaranteed to work for large enough $z$, depending on the 
order of the expansion. In the original work~\cite{KorTrick1}, one has to analytically continue down to $z=1/2$ which appears to be dangerous. 
Nevertheless, the machinery works well at one-loop. We shall find surprises at two loops.

\medskip
As a second comment, we notice that the Baxter equation is invariant under any transformation 
\be
Q(u)\to h(u)\,Q(u),\qquad\mbox{with}\qquad h(u)=h(u+i).
\ee
In other word, $F(z)$ is determined modulo an additive term with periodicity 1 in the $z$ variable. In the above one-loop examples,
this term is absent. Heuristically, this can be explained as follows. Given a polynomial $Q(u)$, it seems reasonable to conclude
that at each order in $1/N$, $\log Q(u)$ must grow at most polynomially with $u$. A periodic contribution in the $z$ variable would instead lead to 
exponentially growing quantities. Certainly, it would be nice to understand this point better.

\medskip
We now move to the more interesting two-loop case where we shall show the partial failure of the $\Delta$-method.

\section{The two-loop Baxter equation for the ground state}

\subsection{General structure}

The long range Baxter equation for the $\mathfrak{sl}(2)$ sector of ${\cal N}=4$ SYM is discussed at three loop accuracy in~\cite{Belitsky:2006wg}.
Here, we shall be interested in the two loop reduction that we discuss in full details in the case of the ground state.

\medskip
We shall require the following definitions
\ba
x(u) &=& \frac{u}{2}\left(1+\sqrt{1-\frac{2\,\lambda}{u^2}}\right), \\
u_\pm &=& u\pm\frac{i}{2}, \qquad x_\pm = x(u_\pm).
\ea
Also, for $\sigma = \pm 1$, we introduce
\be
\Lambda_\sigma = \left.\frac{d}{du}\log\,Q(u)\right|_{u = \frac{i}{2}\,\sigma},\qquad
\Delta_\sigma(x) = x^L\,\exp\left(-\frac{\lambda}{x}\,\Lambda_\sigma\right).
\ee
Notice that $\Lambda_\sigma$ is independent on the spectral parameter $u$. However, it depends on the coupling hidden in the 
Baxter function $Q(u)$. 

\medskip
The long-range Baxter equation reads
\be
\Delta_+(x_+)\,Q(u+i)+\Delta_-\left(x_-\right)\,Q(u-i) = t_L(u)\,Q(u),
\ee
where $t_L(u)$ gets radiative corrections to the conserved charges. For the twist-2 and twist-3 ground state, 
this simply means
\ba
t_2(u) &=& 2\,u^2 + q_{2,2}^{(0)} + \lambda\,q_{2,2}^{(1)}, \\
t_3(u) &=& 2\,u^3 + u\,\left(q_{3,2}^{(0)} + \lambda\,q_{3,2}^{(1)}\right),
\ea
where $q_{L,2}^{(0)}$ is the one-loop charge and $q_{L,2}^{(1)}$ a two loop correction to be computed.

\medskip
In the ground state, this equation must be solved in terms of 
\be
Q(u) = Q^{(0)}(u) + \lambda\,Q^{(1)}(u),
\ee
where the polynomials $Q^{(0)}$, $Q^{(1)}$ are {\em even} and with degrees $N$, $N-2$ respectively. Of course $Q^{(0)}$ is the one-loop
Baxter polynomial. 
Given $Q(u)$, the two-loop anomalous dimension is 
\be
\gamma = \gamma_1\,\lambda + \gamma_2\,\lambda^2 = 
2\,i\,\lambda\,\left[\left. (\log Q)'\right|_{u=i/2} + \frac\lambda 4\,\left. (\log Q)'''\right|_{u=i/2}\right].
\ee
It is convenient to write
\be
\log\,Q(u) = F^{(0)}(u) + \lambda\,F^{(1)}(u) + {\cal O}(\lambda^2).
\ee
Then, an alternative expression for $\gamma$ is 
\be
\gamma =  \left. 2\,i\,\lambda\,\left[ {F^{(0)}}' + \lambda \,\left( {F^{(1)}}'+\frac{1}{4}\,{F^{(0)}}'''\right)\right] \right|_{u=i/2}.
\ee

To go on, we need the two-loop charge $q_{L,2}^{(1)}$. This is easy to compute. For both $L=2,3$ we have
\be
\Lambda_\pm = \mp \frac{i}{2}\,\gamma_1,
\ee
where $\gamma_1 = \gamma_1(N)$ is the one-loop anomalous dimension. Expanding $\Delta_\pm(x_\pm)$, we find
\ba
L=2, \quad \Delta_\pm(x_\pm) &=& u_\pm^2 + \lambda\,\left(-1-u_\pm\,\Lambda_\pm\right) + {\cal O}(\lambda^2) \\
L=3, \quad \Delta_\pm(x_\pm) &=& u_\pm^3 + \lambda\,u_\pm\,\left(-\frac{3}{2}-u_\pm\,\Lambda_\pm\right) + {\cal O}(\lambda^2).
\ea
We replace these expansions in the Baxter equation and match the leading terms powers of $u$. After a short calculation, we obtain 
the following compact results
\ba
L=2, \quad q_{2, 2}^{(1)} &=& -2-\frac{1}{2}\,(1+2\,N)\,\gamma_1, \\
L=3, \quad q_{3, 2}^{(1)} &=& -3-(1+N)\,\gamma_1.
\ea
To summarize, at two-loop order, we want to study the large $N$ expansion of the following truncated Baxter equations

\medskip
{\bf twist-2}
$$
\left[u_+^2 + \lambda\left(\frac{i}{2}\,\gamma_1\,u_+-1\right)\right]\,Q(u+i) + 
\left[u_-^2 + \lambda\left(-\frac{i}{2}\,\gamma_1\,u_--1\right)\right]\,Q(u-i) = 
$$
\be
= \left(2\,u^2+q_{2,2}^{(0)}+\lambda\,q_{2,2}^{(1)}\right)\, Q(u),
\ee

{\bf twist-3}
$$
\left[u_+^3 + \lambda\,u_+\left(\frac{i}{2}\,\gamma_1\,u_+-\frac{3}{4}\right)\right]\,Q(u+i) + 
\left[u_-^3 + \lambda\,u_-\left(-\frac{i}{2}\,\gamma_1\,u_--\frac{3}{4}\right)\right]\,Q(u-i) = 
$$
\be
= u\,\left(2\,u^2+q_{3,2}^{(0)}+\lambda\,q_{3,2}^{(1)}\right)\, Q(u)~~~.
\ee

\bigskip
These equations are rather complicated but indeed admit the desired polynomial solutions. 
As a concrete example, we list here their solutions at $N=4$ 

\medskip

{\bf twist-2}
\ba
Q(u) &=& \left(\frac{35 u^4}{12}-\frac{65 u^2}{24}+\frac{9}{64}\right)+\left(\frac{25}{14}-\frac{80 u^2}{7}\right) \lambda +{\cal O}(\lambda^2), \\
\gamma &=& \frac{25 \lambda }{3}-\frac{925 \lambda ^2}{54}+O\left(\lambda ^3\right).
\ea

{\bf twist-3}
\ba
Q(u) &=& \left(\frac{3 u^4}{2}-\frac{9 u^2}{4}+\frac{11}{32}\right)+\left(\frac{47}{16}-\frac{27 u^2}{4}\right) \lambda +O\left(\lambda ^2\right),\\
\gamma &=& 6 \lambda -\frac{39 \lambda ^2}{4}+O\left(\lambda ^3\right).
\ea

\subsection{Two-loop conjectures for $\gamma$ and their large $N$ expansions}

In literature, we can find two loop conjectures for $\gamma$ in twist-2 and twist-3. In the next sections, we shall 
report their expressions as well as the rigorous large $N$ expansions.

\subsubsection{Twist-2}

The two-loop conjecture for $\gamma_2$ in twist-2 and even $N$ is well-known~\cite{Kotikov:2004er}. We adhere to the notation of~\cite{Beccaria:2007cn}.
It reads 
\ba
\gamma_2 &=& -4\,\left[S_3(N)+S_{-3}(N)-2\,S_{-2,1}(N)+2\,S_1(N)\,(S_2(N)+S_{-2}(N))\right] = \nonumber \\ \nonumber\\
&=& -S_3\left(\frac{N}{2}\right)+8\,S_{-2,1}(N)-4\,S_1(N)\,S_2\left(\frac{N}{2}\right).
\ea
The large $N$ expansion of this expression can be derived with minor effort. The simple harmonic sums $S_a(N)$ or $S_a(N/2)$ 
can be expanded by using the known result 
\be
S_a(N) = \zeta_a + \frac{a-2\,N-1}{2\,(a-1)\,N^a}-\frac{1}{(a-1)!}\sum_{k\ge 1}\frac{(2\,k+a-2)!\,B_{2\,k}}{(2\,k)!\,N^{2\,k+a-1}}, \qquad a\in \mathbb{N}, a>1.
\ee
The nested sum $S_{-2,1}(N)$ is a bit tricky. A convenient heuristic procedure is as follows. We start from 
the recurrence relation
\be
S_{-2,1}(N+2)-S_{-2,1}(N) = \frac{S_1(N+2)}{(N+2)^2}-\frac{S_1(N+1)}{(N+1)^2},\quad N\,\mbox{even},
\ee
and insert in place of $S_{-2,1}(N)$ a generic logarithmic expansion for large $N$. The various terms $\sim\log^n N/N^m$ can be easily matched and the final result is 
\ba
S_{-2,1}\left(\frac{1}{\varepsilon}\right) &=& -\frac{5 \zeta_3}{8}+\left(\frac{\gamma_E }{2}-\frac{\log\,\varepsilon}{2}\right) \varepsilon ^2+\left(\frac{\log\,\varepsilon}{2}-\frac{\gamma_E
   }{2}+\frac{1}{2}\right) \varepsilon ^3-\frac{5 \varepsilon ^4}{12}+\nonumber \\
&& + \left(-\frac{\log\,\varepsilon}{2}+\frac{\gamma_E }{2}-\frac{11}{24}\right) \varepsilon ^5+\frac{151
   \varepsilon ^6}{240}+\left(\frac{3 \log\,\varepsilon}{2}-\frac{3 \gamma_E }{2}+\frac{469}{240}\right) \varepsilon ^7 + \nonumber\\
&& -\frac{331 \varepsilon ^8}{126}+\left(-\frac{17 
\log\,\varepsilon}{2}+\frac{17 \gamma_E }{2}-\frac{67379}{5040}\right) \varepsilon ^9+\cdots~~~.
\ea
Combining these partial expansions, we obtain the desired expansion of $\gamma_2$ as
\ba
\label{eq:gamma2exptwist2}
\gamma_2 &=& \left(\frac{2}{3} \pi ^2 \log\,\overline{\varepsilon}-6 \zeta_3\right)+\left(-8 \log\,\overline{\varepsilon}-\frac{\pi ^2}{3}\right) \varepsilon +
\left(4 \log\,\overline{\varepsilon}+\frac{\pi
   ^2}{18}+6\right) \varepsilon ^2+\\
&& + \left(-\frac{4 \log\,\overline{\varepsilon}}{3}-\frac{14}{3}\right) \varepsilon ^3+\left(4-\frac{\pi ^2}{180}\right) \varepsilon ^4+\left(\frac{4
   \log\,\overline{\varepsilon}}{15}-\frac{182}{45}\right) \varepsilon ^5+\left(-\frac{5}{2}+\frac{\pi ^2}{378}\right) \varepsilon ^6+\cdots ,\nonumber
\ea
where $\overline\varepsilon = e^{-\gamma_E}\,\varepsilon$. In the following, we shall {\em systematically omit the terms proportional to $\gamma_E$} since they can 
all be generated by the replacement $\log\varepsilon\to\log\overline\varepsilon$. In the above expressions, we have defined
\be
\zeta_n = \zeta(n) = \sum_{k=1}^\infty \frac{1}{k^n}.
\ee

\subsubsection{Twist-3}

The two-loop conjecture for $\gamma_2$ in twist-3 has been obtained independently in~\cite{Beccaria:2007cn,Kotikov:2007cy}
and takes the simple form 
\be
\gamma_2 = -2\,S_3\left(\frac{N}{2}\right)-4\,S_1\left(\frac{N}{2}\right)\,S_2\left(\frac{N}{2}\right).
\ee
The large $N$ expansion can be evaluated without particular difficulties and it finally reads
\ba
\label{eq:gamma2exptwist3}
\gamma_2 &=& \left(\frac{2}{3} \pi ^2 \log\,\overline{\varepsilon}-2 \zeta_3\right)+\left(-4 \log\,\overline{\varepsilon}-\frac{\pi ^2}{3}\right) 
\varepsilon +\left(2 \log\,\overline{\varepsilon}+\frac{\pi
   ^2}{18}+3\right) \varepsilon ^2+\left(-\frac{2 \log\,\overline{\varepsilon}}{3}-\frac{7}{3}\right) \varepsilon ^3+\nonumber\\
&& + \left(1-\frac{\pi ^2}{180}\right) \varepsilon ^4+\left(\frac{2
   \log\,\overline{\varepsilon}}{15}-\frac{1}{45}\right) \varepsilon ^5+\left(-\frac{1}{4}+\frac{\pi ^2}{378}\right) \varepsilon ^6+\cdots . 
\ea

\subsection{Large $N$ expansion from the Baxter equation: $\Delta$-method in twist-2}

We repeat the same kind of analysis we did in the one-loop case. The two loop Baxter equation involves $S_1(N)$ which we expand to any desired
order in $\varepsilon$. Writing
\ba
F(z) &=& F^{(0)}(z)+\lambda\,F^{(1)}(z), \\
\Delta(z) &=& F(z+1)-F(z) =  \Delta^{(0)}(z) + \lambda\,\Delta^{(1)}(z),
\ea
we find that the new contribution $\Delta^{(1)}(z)$ can be plainly matched to the equation if it takes the general form (
we recall again that in the following we set $\varepsilon\equiv \overline\varepsilon$ for simplicity)
\be
\Delta^{(1)}(z) = \sum_{n=0}^\infty (a_n(z)\,\log\varepsilon+b_n(z))\,\varepsilon^n. 
\ee
The logarithmic enhancement is expected given the exact large $N$ expansion of $\gamma_2$. Besides, it is required from the equation after expansion.
The actual expression of $\Delta^{(1)}(z)$ reads
\ba
\Delta^{(1)}(z) &=& \left(\frac{4 \log\,\varepsilon}{2 z+1}-\frac{4}{(2 z+1)^2}\right)+\left(-4 \log\,\varepsilon-\frac{2}{2 z+1}\right) \varepsilon +\\
&& + \left(2 \log\,\varepsilon+\frac{1}{3 (2
   z+1)}+4\right) \varepsilon ^2 +\left(8 z^2 \log\,\varepsilon-\frac{10}{3}\right) \varepsilon ^3+\nonumber\\
&& + \left(-10 z^2+2 z+\left(-12 z^2-1\right) \log\,\varepsilon-\frac{1}{30 (2 z+1)}+\frac{2}{3}\right) \varepsilon ^4+\cdots . \nonumber
\ea
The expansion can apparently be continued to any order and no special problems do appear. However, if we compute the predicted $\gamma_2$, we encounter a 
subtle problem already at order $\varepsilon^4$. The anomalous dimension 
is given by 
\be
\gamma_2 = 2\, \lim_{z\to \frac{1}{2}} \left[(F^{(1)}(z))'-\frac{1}{4}\,(F^{(0)}(z))'''\right] .
\ee
Evaluating $F^{(1)}(z)$ by the same methods we used at one-loop and solving the difference equations associated with $\Delta^{(1)}(z)$, we easily find
\ba
\gamma_{2, \Delta-{\rm method}} &=& \left(\frac{2}{3} \pi ^2 \log\,\overline{\varepsilon}-6 \zeta_3\right)+\left(-8 \log\,\overline{\varepsilon}-\frac{\pi ^2}{3}\right) \varepsilon +
\left(4 \log\,\overline{\varepsilon}+\frac{\pi
   ^2}{18}+6\right) \varepsilon ^2+\nonumber\\
&& + \left(-\frac{4 \log\,\overline{\varepsilon}}{3}-\frac{14}{3}\right) \varepsilon ^3+\left(2-\frac{\pi ^2}{180}\right) \varepsilon ^4+\cdots~~~.
\ea
Comparing with the exact result \refeq{eq:gamma2exptwist2}, one sees that a mismatch appears at order $\varepsilon^4$. 
Expanding the calculation to higher orders, one finds
\be
\label{eq:twist2mismatch}
\gamma_{2, \Delta-{\rm method}}-\gamma_2 = -2\, \varepsilon ^4+4\, \varepsilon ^5+2\, \varepsilon ^6-16\, \varepsilon ^7-6\, \varepsilon ^8+108\, \varepsilon ^9+34\, \varepsilon ^{10}
+O\left(\varepsilon ^{11}\right).
\ee
All terms of the above mismatch are not transcendental neither have logarithmic enhancement. So, we conclude that 
the $\Delta$-method works at leading logarithmic accuracy including also 
the non-enhanced transcendental terms. However, it does not work for the rational contributions starting from ${\cal O}(\varepsilon^4)$~!

This failure seems to be related to the fact that the two-loop terms $\sim u_-$ in the l.h.s. of the Baxter equations 
do not vanish at $z=1/2$. Later, we shall clarify this point.

\subsection{Large $N$ expansion from the Baxter equation: $\Delta$-method in twist-3}

The twist-3 case is much more easy because the special structure of the two-loop Baxter equation. 
The terms $\sim u_-$ in the l.h.s. of the Baxter equations vanish at $z=1/2$. A straightforward calculation 
provides the following expression of the two-loop difference function $\Delta^{(1)}(z)$ 
\ba
\Delta^{(1)}(z) &=& \frac{2 \log\,\varepsilon}{z+\frac{1}{2}}-\frac{6}{(2 z+1)^2}+\varepsilon  \left(-2 \log\,\varepsilon-\frac{2}{2 z+1}\right)+ \nonumber\\
&& \varepsilon ^2 \left(\frac{42 z+25}{12 (2 z+1)}+\log\,\varepsilon\right) +\varepsilon ^3 \left(\frac{1}{4} \left(4 z^2-1\right) \log\,\varepsilon-\frac{17}{12}\right)+\nonumber \\
&&\varepsilon ^4 \left(\frac{-4080 z^2+720 z-\frac{45}{z-1}-\frac{128}{2 z+1}+2240+\frac{45}{z}}{3840}+\frac{1}{8} \left(-12 z^2-1\right) \log\,\varepsilon\right)+ \nonumber \\
&& \varepsilon ^5 \left(\frac{47 z^2}{24}-\frac{3 z}{8}+\frac{3}{128} \left(\frac{1}{z-1}-\frac{1}{z}\right)+\frac{7}{120}+\right. \nonumber \\
&&  \left. \left(-\frac{3 z^4}{4}+\frac{z^3}{2}+\frac{13 z^2}{16}+\frac{3 z}{16}+\frac{1}{256} \left(\frac{1}{z}-\frac{1}{z-1}\right)+\frac{11}{64}\right) 
\log\,\varepsilon\right)+\cdots\nonumber .
\ea
The contributions from the rational functions can be evaluated in $F'(1/2)$ by using in particular the special values
\be
\begin{array}{lll}
\psi(1/2) &=& -\gamma_E-2\,\log\,2, \\
\psi'(1/2) &=& \pi^2/2, \\
\psi''(1/2) &=& -14\,\zeta_3, \\
\psi'''(1/2) &=& \pi^4, 
\end{array}
\qquad
\begin{array}{lll}
\psi(-1/2) &=& 2-\gamma_E-2\,\log\,2, \\
\psi'(-1/2) &=& 4+\pi^2/2, \\
\psi''(-1/2) &=& 16-14\,\zeta_3, \\
\psi'''(-1/2) &=& 96+\pi^4. 
\end{array}
\ee
A short calculation shows that the correct complete two loop large spin expansion of the anomalous dimension \refeq{eq:gamma2exptwist3}
is perfectly reproduced by starting from the above expression.

\section{The improved expansion of the Baxter equation in twist-2}

Let us go back to the weak points of the $\Delta$-method. We are assuming an expansion for $\Delta(z)$ valid in the Baxter equation for 
both $\Delta(z)$ and $\Delta(z-1)$ in a neighborhood of $z=1/2$. The assumed expansion is clearly wrong as it stands. For instance, by parity invariance
we have rigorously $\Delta(-1/2) = 0$ which is not obvious in the expansion derived at large $z>0$.

To see what is happening, we look at the leading term in the $\varepsilon\to 0$ expansion of $F(z)$. It is easily derived as 
\be
F(z) = -2\,|z|\,\log\varepsilon + \cdots~~~.
\ee
If $z>1$, we obtain 
\ba
\Delta(z) &=& -2\,\log\varepsilon + \cdots, \\
-\Delta(z-1) &=& +2\,\log\varepsilon + \cdots~~~.
\ea
The different signs are responsible for the suppression of the second piece of the Baxter equation
\be
e^{\Delta(z)}\sim \frac{1}{\varepsilon^2},\qquad
e^{-\Delta(z-1)}\sim \varepsilon^2.
\ee
However, if $z$ is around $1/2$, the value we are interested in, we find, for small enough $\rho$
\ba
z &=& \frac{1}{2}+\rho, \\
\Delta(z) &=& -2\,\log\varepsilon + \cdots, \\
-\Delta(z-1) &=& -F(z)+F(z-1) = 2\,\left(\left|\frac{1}{2}+\rho\right|-\left|-\frac{1}{2}+\rho\right|\right)\,\log\varepsilon + \cdots = \nonumber \\
&=& 4\,\rho\,\log\varepsilon + \cdots~~~.
\ea
These expansions lead to the completely altered balance of the two terms in the l.h.s. of the Baxter equation
\be
e^{\Delta(z)}\sim \frac{1}{\varepsilon^2},\qquad
e^{-\Delta(z-1)}\sim \frac{1}{\varepsilon^2}\,\varepsilon^{4\,z}.
\ee
In particular, this means that the expansion of $F$ will contain powers of $\varepsilon^{4\,z}$ multiplied by eventual 
additional integer powers of $\varepsilon$ and possible logarithmic enhancements.

We shall see that these {\em anomalous terms} are actually present and are highly non-trivial. At one-loop they will not 
contribute in both twist 2 and 3 giving a rigorous support to the applicability of the $\Delta$-method to the one-loop Baxter equation. 
However, at two-loops they will give a non-trivial crucial contribution in twist-2, being still negligible in twist-3.
Later, we shall show that they must be included in the three loop analysis of twist-3. In conclusion, a safe procedure amounts to compute them
to see if they are relevant or negligible.

Now, let us go back to the solution of the twist-2 problem. We have seen that 
we have to work with the function $F(z)$ forced to be even under $z\to -z$ and including in its 
asymptotic expansion possible anomalous terms according to 
\ba
\label{eq:genexp}
F^{(0)}(z) &=& -2\,z\,\log\varepsilon+\sum_{n=0}^\infty a_n^{(0)}(z)\,\varepsilon^n + \sum_{m=1}^\infty\sum_{n=0}^\infty f_{m,n}^{(0)}(z)\,\varepsilon^{4\,m\,z+n}, \\
F^{(1)}(z) &=& \sum_{n=0}^\infty (a_n^{(1)}(z)+b_n^{(1)}(z)\,\log\varepsilon)\,\varepsilon^n + \sum_{m=1}^\infty\sum_{n=0}^\infty (f_{m,n}^{(1)}(z)+
g_{m,n}^{(1)}(z)\,\log\varepsilon)\,\varepsilon^{4\,m\,z+n}. \nonumber
\ea
Notice that, in principle, the anomalous pieces give contributions to both the logarithmic and non-logarithmic
terms of the anomalous dimension. As a check, we shall also observe the cancellation of the logarithmic extra contributions.

Very remarkably, the non-anomalous functions $a_n^{(0)}$, $a_n^{(1)}$ and $b_n^{(1)}$ turns out to be equal to those computed in the $\Delta$-method.
For completeness, we shall report them, since we have already given their contribution to $\Delta(z)$, but not to $F(z)$.
Instead, the anomalous functions $f_{m, n}^{(0)}$, $f_{m, n}^{(1)}$ and $g_{m, n}^{(1)}$ are very non-trivial.

Here are the explicit results that can be found with some labor expanding the Baxter equation. Of course, one cannot use anymore the difference 
function $\Delta(z)$, but use instead the correct replacements valid in a neighborhood of $z=1/2$
\ba
\Delta(z) &=& F(z+1)-F(z), \\
\Delta(z-1) &=& F(z)-F(z-1) = F(z)-F(1-z).
\ea
After these replacements, we insert in the Baxter equation the generalized asymptotic expansion of $F(z)$ for $z>0$ described in \refeq{eq:genexp}.

We give all the expressions that are needed to 
reproduce the anomalous dimension at order ${\cal O}(\varepsilon^5\,\log\varepsilon)$ included. This requires to consider the following functions
\ba
a_n^{(\ell)} &,& 0\le n \le 5, \quad \ell = 0, 1\nonumber \\
f_{1,n}^{(\ell)} &,& 0\le n\le 3, \quad \ell = 0, 1, \nonumber \\
b_n^{(1)} &,& 0\le n \le 5, \\
g_{1,n}^{(1)} &,& 0\le n\le 3, \nonumber \\
f_{2,n}^{(\ell)} &,& n = 0, 1, \quad \ell = 0, 1, \nonumber \\
g_{2,n}^{(1)} &,& n = 0, 1. \nonumber
\ea

First the non-anomalous one-loop contributions
\ba
a_0^{(0)}(z) &=& -2\,\log\,\Gamma\left(z+\frac{1}{2}\right), \\
a_1^{(0)}(z) &=& z, \\
a_2^{(0)}(z) &=& \frac{1}{3}\,z\,(z-1)\,(2\,z-1), \\
a_3^{(0)}(z) &=& -\frac{1}{6}\,z\,(3-6\,z+4\,z^2), \\
a_4^{(0)}(z) &=& -\frac{1}{240} \,z\, (2\, z-1) \,\left(72\, z^3-204\, z^2+158\, z-41\right), \\
a_5^{(0)}(z) &=& \frac{1}{40} \,z\, \left(48\, z^4-160\, z^3+200\, z^2-120\, z+35\right)
\ea
Then, the two-loop non-anomalous contributions
\be
\begin{array}{lll}
a_0^{(1)} &=& \psi^{(1)}\left(z+\frac{1}{2}\right), \\
a_1^{(1)} &=& -\psi\left(z+\frac{1}{2}\right), \\
a_2^{(1)} &=& 4\,z + \frac{1}{6}\,\psi\left(z+\frac{1}{2}\right), \\
a_3^{(1)} &=& -\frac{10}{3}\,z, \\
a_4^{(1)} &=& -2\,z+6\,z^2-\frac{10}{3}\,z^3-\frac{1}{60}\,\psi(\left(z+\frac{1}{2}\right)), \\
a_5^{(1)} &=& \frac{1}{45} \,z\, \left(280\, z^2-510\, z+299\right), \\
\end{array}\qquad
\begin{array}{lll}
b_0^{(1)} &=& 2\,\psi\left(z+\frac{1}{2}\right), \\
b_1^{(1)} &=& -4\,z, \\
b_2^{(1)} &=& 2\,z, \\
b_3^{(1)} &=& \frac{4}{3}\,z\,(z-1)\,(2\,z-1), \\
b_4^{(1)} &=& -z\,(3-6\,z+4\,z^2), \\
b_5^{(1)} &=& -\frac{1}{30}\, z\, (2\, z-1) \times \\
&& \,\left(72 \,z^3-204\, z^2+158\,  z-41\right) .
\end{array}
\ee
Now, the more interesting anomalous terms. We begin with those $\sim \varepsilon^{4\,z+n}$. The one-loop contributions are 
\be
\begin{array}{lll}
\displaystyle f_{1,0}^{(0)}(z) &=& \displaystyle\frac{\Gamma\left(\frac{1}{2}+z\right)^2}{\Gamma\left(\frac{1}{2}-z\right)^2}, \\ \\
f_{1,1}^{(0)}(z) &=& \displaystyle-2\,z\,\frac{\Gamma\left(\frac{1}{2}+z\right)^2}{\Gamma\left(\frac{1}{2}-z\right)^2}, 
\end{array}
\qquad
\begin{array}{lll}
f_{1,2}^{(0)}(z) &=& \displaystyle-\frac{2\,\pi^2\,z\,(z-1)\,(2\,z-1)}{3\,\cos^2(\pi\,z)\,\Gamma\left(\frac{1}{2}-z\right)^4},\\ \\
f_{1,3}^{(0)}(z) &=& \displaystyle\frac{\pi^2\,z\,(3+4\,z+8\,z^3)}{3\,\cos^2(\pi\,z)\,\Gamma\left(\frac{1}{2}-z\right)^4}.
\end{array}
\ee
Each of these functions has vanishing derivative in $z=1/2$. This means that no anomalous contributions appear  at one-loop, 
recovering the correctness of the $\Delta$-method at one loop. However, two-loop non trivial contributions are associated with the third derivative.
For instance
\be
\lim_{z\to\frac{1}{2}} \frac{d^3}{dz^3}\left[f_{1,0}^{(0)}(z)\,\varepsilon^{4\,z}\right] = 24\,\varepsilon^2\,\log\overline\varepsilon.
\ee

The two-loop anomalous pieces are rather complicated. Let us define
\be
D(z) = \psi^{(1)}\left(\frac{3}{2}+z\right)-\psi^{(1)}\left(\frac{3}{2}-z\right).
\ee
We have 
\ba
f_{1,0}^{(1)}(z) &=& -\frac{\pi^2\,(-32\,z+D(z)\,(1-4\,z^2)^2)}{(1+2\,z)^4\,\cos^2(\pi\,z)\,\Gamma\left(-\frac{1}{2}-z\right)^2\,\Gamma\left(\frac{3}{2}-z\right)^2}, \\
\nonumber \\
f_{1,1}^{(1)}(z) &=& \frac{\pi^2\,\sec^2(\pi\,z)\,(2\,z\,(-32\,z+D(z)\,(1-4\,z^2)^2))+\pi\,(1-4\,z^2)^2\,\tan^2(\pi\,z)}{(1+2\,z)^4\,\,
\Gamma\left(-\frac{1}{2}-z\right)^2\,\Gamma\left(\frac{3}{2}-z\right)^2}, \nonumber 
\ea
and
\ba
f_{1,2}^{(1)}(z) &=& \frac{M_2}{6 (2 z-1) (2 z+1)^2 \Gamma \left(\frac{1}{2}-z\right)^4}, \\
f_{1,3}^{(1)}(z) &=& \frac{M_3}{3 (1-2 z)^2 (2 z+1) \Gamma \left(\frac{1}{2}-z\right)^4}
\ea
with 
\ba
M_2 &=& \pi ^2 \sec ^2(\pi  z) \left(4 z \left(D(z) (z-1) \left(1-4 z^2\right)^2-8 z (2 z (6 z+5)-7)+12\right) +\right. \nonumber\\
&& \left. -\pi  (2 z-1) (2 z+1)^2 (12 z+1) \tan (\pi    z)\right), \\
\nonumber\\
M_3  &=& \pi ^2 z \sec ^2(\pi  z) \left(-\pi  (2 z+1) \left(4 z^2-6 z+1\right) \tan (\pi  z) (1-2 z)^2 + \right. \nonumber\\
&& \left. -D(z)\left(1-4 z^2\right)^2 \left(4 z^2-2 z+3\right)+4
   \left(2 z \left(6 z \left(8 z^2+2 z-5\right)+13\right)+5\right)\right) \nonumber
\ea
Finally, we have 
\ba
g_{1,0}^{(1)}(z) &=& -\frac{2 \pi ^3 \tan (\pi  z)}{\cos^2(\pi\,z)\,\Gamma \left(\frac{1}{2}-z\right)^4},\\
g_{1,1}^{(1)}(z) &=& \frac{4 \pi ^2 z (\pi  \tan (\pi  z)+2)}{\cos^2(\pi\,z)\,\Gamma \left(\frac{1}{2}-z\right)^4}, \\
g_{1,2}^{(1)}(z) &=& \frac{4 \pi ^2 z  (-12 z+\pi  (z-1) (2 z-1) \tan (\pi  z)-3)}{3\, \cos^2(\pi\,z)\,\Gamma \left(\frac{1}{2}-z\right)^4}, \\
g_{1,3}^{(1)}(z) &=& -\frac{2 \pi ^2 z (2 z+1)  \left(8 z^2-12 z+\pi  \left(4 z^2-2 z+3\right) \tan (\pi  z)+4\right)}
{3\,\cos^2(\pi\,z)\, \Gamma \left(\frac{1}{2}-z\right)^4}
\ea

The anomalous contributions $\sim \varepsilon^{8\,z+n}$ can also be computed, however we have checked that in all cases they do not give contributions to the
anomalous dimension. This is due to the fact that the one loop contributions are ${\cal O}((z-1/2)^4)$ while the two loop contributions are ${\cal O}((z-1/2)^2)$.
Some examples at one-loop are 
\ba
f^{(0)}_{2,0}(z) &=& \frac{(2 z+1)^2 \Gamma \left(z+\frac{1}{2}\right)^4-8 \Gamma \left(z+\frac{1}{2}\right)^2 \Gamma \left(z+\frac{3}{2}\right)^2}{2 (2 z+1)^2 \Gamma
   \left(\frac{1}{2}-z\right)^4}= \\
&=& -\frac{1}{2} \left(z-\frac{1}{2}\right)^4+{\cal O}\left(\left(z-\frac{1}{2}\right)^5\right) \\
f^{(0)}_{2,1}(z) &=& 
\frac{2 \pi ^2 z \Gamma \left(z+\frac{1}{2}\right)^2}{\cos^2(\pi z)\,\Gamma \left(\frac{1}{2}-z\right)^6} = \left(z-\frac{1}{2}\right)^4+{\cal O}\left(\left(z-\frac{1}{2}\right)^5\right)
\ea
A two loop example is 
\ba
f^{(1)}_{2,0}(z) &=& \frac{\pi ^4 \left(-\psi ^{(1)}\left(\frac{3}{2}-z\right) \left(1-4 z^2\right)^2+\psi ^{(1)}\left(z+\frac{3}{2}\right) \left(1-4 z^2\right)^2-32 z\right)}
{\cos^4(\pi\,z)\,\left(1-4 z^2\right)^2 \Gamma \left(\frac{1}{2}-z\right)^8} = \nonumber \\
&=& -\left(z-\frac{1}{2}\right)^2+{\cal O}\left(\left(z-\frac{1}{2}\right)^3\right) \\
g^{(1)}_{2,0}(z) &=& \frac{2 \pi ^5 \tan (\pi  z)}{\cos^4(\pi  z) \Gamma \left(\frac{1}{2}-z\right)^8} = -2 \left(z-\frac{1}{2}\right)^3+{\cal O}\left(\left(z-\frac{1}{2}\right)^4\right)
\ea

Collecting all results, we find the following contributions to the two loop anomalous dimension. From the one-loop anomalous pieces
\be
\left. (F_{\rm anom}^{(0)}(z))'''\right|_{z=1/2} = 
24 \,\varepsilon^2\,\log\,\varepsilon -12 \,(2\, \log\,\varepsilon+1) \varepsilon ^3+2\, \varepsilon ^4+2\, (12\, \log\,\varepsilon+11) \varepsilon ^5+\cdots
\ee
From the two-loop anomalous pieces
\be
\left. (F_{\rm anom}^{(1)}(z))'\right|_{z=1/2} = 
6 \,\varepsilon^2\,\log\,\varepsilon-3\, (2\, \log\,\varepsilon+1) \varepsilon ^3+\frac{3 \,\varepsilon ^4}{2}+\frac{1}{2}\, (12\, \log\,\varepsilon+7) \varepsilon ^5+\cdots
\ee
The full {\em anomalous} contribution is the combination 
\be
\left. 2\,\left[
(F_{\rm anom}^{(1)}(z))'-\frac{1}{4}(F_{\rm anom}^{(0)}(z))'''\right]\right|_{z=1/2} = 2\,\varepsilon^4-4\,\varepsilon^5 + \cdots
\ee
which is precisely the required piece to correct the mismatch in \refeq{eq:twist2mismatch} to order ${\cal O}(\varepsilon^5\,\log\varepsilon)$.
We have extended the calculation to higher orders. The whole procedure is easily automatized and in all cases, the mismatch is corrected.

\section{The three loop Baxter equation: Twist-3}

To conclude, we now analyze the three loop Baxter equation in twist-3 to show that even in this case, it is necessary to include
anomalous terms to match the rational part of the large spin anomalous dimension.

We have to extend the notation we introduced for the two loop case. To this aim, we define
for $\sigma = \pm 1$
\ba
\Lambda_\sigma^{(n)} &=& \left.\frac{d^n}{du^n}\log\,Q(u)\right|_{u = \frac{i}{2}\,\sigma},\\
\Delta_\sigma(x) &=& x^3\,\exp\left(-\frac{\lambda}{x}\,\Lambda_\sigma^{(1)}-\frac{\lambda^2}{4\,x^2}(\Lambda_\sigma^{(2)}+x\,\Lambda_\sigma^{(3)})\right).
\ea
The Baxter equation reads again
\be
\Delta_+(x_+)\,Q(u+i)+\Delta_-\left(x_-\right)\,Q(u-i) = t_L(u)\,Q(u).
\ee
In the ground state the transfer matrix gets radiative corrections to the unique non-trivial charge
\be
t_3(u) = u\,\left[2\,u^2 + \left(q_{3,2}^{(0)} + \lambda\,q_{3,2}^{(1)} + \lambda^2\,q_{3,2}^{(2)}\right)\right].
\ee
This equation must be solved in terms of 
\be
Q(u) = Q^{(0)}(u) + \lambda\,Q^{(1)}(u) + \lambda^2\,Q^{(2)}(u),
\ee
where the polynomials $Q^{(0)}$, $Q^{(1)}$, $Q^{(2)}$ are even and with degrees $N$, $N-2$, $N-2$ respectively. The 3-loop anomalous dimension
is conveniently expressed in terms of 
\be
F = \log\,Q = F^{(0)} + \lambda\,F^{(1)} + \lambda^2\, F^{(2)} + {\cal O}(\lambda^3) 
\ee
as
\ba
\gamma &=& \gamma_1\,\lambda + \gamma_2\,\lambda^2 +\gamma_3\,\lambda^3 = \nonumber \\
&=& 2\,i\,\left[{F^{(0)}}'\,\lambda + \left({F^{(1)}}'+\frac{1}{4}\,{F^{(0)}}'''\right)\,\lambda^2 + \left({F^{(2)}}'+\frac{1}{4}\,
{F^{(1)}}'''+\frac{1}{48}\,{F^{(0)}}'''''\right)\,\lambda^3\right],
\ea
where all derivatives are evaluated at $u=i/2$.

A tedious but straightforward calculation gives the three loop expansion of $\Delta_\pm(x_\pm)$
in terms of the one and two-loop anomalous dimensions 
\be
\Delta_\pm(x_\pm) = u_\pm^3 + u_\pm\,\left(-\frac{3}{2}\pm\frac{i}{2}\,u_\pm\,\gamma_1\right)\,\lambda + \left(\mp\frac{i}{2}\,\gamma_1
-\frac{1}{8}\,u_\pm\,\gamma_1^2\pm\frac{i}{2}\,u_\pm^2\,\gamma_2\right)\,\lambda^2 + \cdots
\ee
as well as the three loop correction to the second charge
\be
q_{3,2}^{(2)} = -\frac{1}{4}\,\gamma_1^2-(1+N)\,\gamma_2.
\ee
Notice that it is remarkable that such simple expressions can be obtained. They are  completely determined by the 
previous calculations at lower orders, {\em i.e.} $\gamma_1$ and $\gamma_2$. Notice also that what is actually required is just their
large $N$ expansion. The resummed closed form is not necessary.

In conclusion, the 3-loop truncated Baxter equation is 
\ba
\left[u_+^3 + u_+\,\left(-\frac{3}{2}+\frac{i}{2}\,u_+\,\gamma_1\right)\,\lambda + \left(-\frac{i}{2}\,\gamma_1
-\frac{1}{8}\,u_+\,\gamma_1^2+\frac{i}{2}\,u_+^2\,\gamma_2\right)\,\lambda^2\right]\,&& Q(u+i) + \nonumber \\
\left[u_-^3 + u_-\,\left(-\frac{3}{2}-\frac{i}{2}\,u_+\,\gamma_1\right)\,\lambda + \left(+\frac{i}{2}\,\gamma_1
-\frac{1}{8}\,u_-\,\gamma_1^2-\frac{i}{2}\,u_-^2\,\gamma_2\right)\,\lambda^2\right]\,&& Q(u-i) =  \nonumber 
\ea
\be
= u\,\left[2\,u^2 + \left(q_{3,2}^{(0)} + \lambda\,q_{3,2}^{(1)} + \lambda^2\,q_{3,2}^{(2)}\right)\right]\,Q(u).
\ee
As we mentioned, the three loop terms in the second term of the l.h.s. do not vanish as $u_-\to 0$. This suggests that anomalous contributions will be present 
at this order.

Again, an example can be useful to check the above equation. For $N=4$ its solution is 
\ba
Q(u) &=& \left(\frac{3 u^4}{2}-\frac{9 u^2}{4}+\frac{11}{32}\right)+\left(\frac{47}{16}-\frac{27 u^2}{4}\right) \lambda +\left(\frac{27 u^2}{8}+\frac{159}{32}\right)
   \lambda ^2+\cdots, \\
\gamma &=& 6 \lambda -\frac{39 \lambda ^2}{4}+\frac{957 \lambda ^3}{32}+\cdots
\ea

\subsection{Three loop conjecture for $\gamma$ and its large $N$ expansion}

The three loop conjecture for $\gamma_3$ is~\cite{Beccaria:2007cn,Kotikov:2007cy}
\be
\gamma_3 = 5\,S_5 + 6\,S_{2}\,S_{3} - 8\,S_{3, 1, 1} + 4\,S_{4, 1} - 4\,S_{2, 3} +  S_{1}\,(4\,S_{2}^2 + 2\,S_{4} + 8\, S_{3, 1})
\ee
with all harmonic sums evaluated at $N/2$.
Its large $N$ expansion can be worked out with no problems and the result is ($2/N = \varepsilon$) 
\ba
\gamma_3 &=& 
-\frac{11}{45} \pi ^4 \log\,\varepsilon+\frac{\pi^2}{3}\,\zeta_3-\zeta_5 +\left(\frac{4}{3} \pi ^2 \log\,\varepsilon-2 \zeta_3
+\frac{11 \pi ^4}{90}\right) \varepsilon + \nonumber \\
&& + \left(-2 \log^2\varepsilon+\left(-4-\frac{2 \pi ^2}{3}\right) \log\,\varepsilon+\zeta_3-\frac{11 \pi
   ^4}{540}-\frac{5 \pi ^2}{6}+1\right) \varepsilon ^2+\nonumber\\
&& + \left(2 \log^2\,\varepsilon+\left(8+\frac{2 \pi ^2}{9}\right) \log\,\varepsilon-\frac{\zeta_3}{3}+\frac{11 \pi
   ^2}{18}+2\right) \varepsilon ^3+\\
&& + \left(-\log^2\,\varepsilon-\frac{23 \log\,\varepsilon}{3}+\frac{11 \pi ^4}{5400}-\frac{\pi ^2}{4}-\frac{125}{24}\right) \varepsilon
   ^4+\nonumber\\
&& + \left(\left(4-\frac{2 \pi ^2}{45}\right) \log\,\varepsilon+\frac{\zeta_3}{15}+\frac{\pi ^2}{135}+\frac{23}{4}\right) \varepsilon ^5+\nonumber\\
&& + \left(\frac{\log^2\,\varepsilon}{3}
+\frac{\log\,\varepsilon}{45}-\frac{11 \pi ^4}{11340}+\frac{\pi ^2}{18}-\frac{448}{135}\right) \varepsilon ^6+\cdots\nonumber .
\ea
Notice that generic double logarithms appear in the expansion. Formal properties of this expression as well as its four loop 
extension are discussed in~\cite{Beccaria:2007bb}.

\subsection{Large $N$ expansion from the Baxter equation: $\Delta$-method}

Following the $\Delta$-method we obtain the following three loop contribution to the function $\Delta(z)$. It is convenient to write it in terms of 
\be
p = z+\frac{1}{2}.
\ee
\ba
\Delta^{(2)}(z) &=& -\frac{\left(\pi ^2 p^2+3\right) \log\,\varepsilon}{3 p^3}+\frac{\zeta_3}{p}+\frac{9}{8 p^4} + \nonumber\\
&& \left[\left(\frac{\pi ^2}{3}+\frac{2}{p}\right) \log\,\varepsilon-\zeta_3+\frac{\pi ^2}{6 p}+\frac{1}{2 p^3}\right]\,\varepsilon + \nonumber\\
&& \left[-\log^2\,\varepsilon + 
\left(-\frac{\pi ^2}{6}-2-\frac{1}{p}\right) \log\,\varepsilon+\frac{\zeta_3}{2}-\frac{\pi ^2}{6}-\frac{\pi ^2}{36 p}-\frac{3}{2 p}-\frac{1}{12 p^3}
\right]\,\varepsilon^2 + \cdots
\ea
We stop at this order since this gives the following three loop anomalous dimension 
\ba
\gamma_{3, \Delta-{\rm method}} &=& \left(-\frac{11}{45} \pi ^4 \log\,\varepsilon-\zeta_5+\frac{1}{3} \pi ^2 \zeta_3\right)+\left(\frac{4}{3} \pi ^2 \log\,\varepsilon-2 \zeta_3+\frac{11 \pi
   ^4}{90}\right) \varepsilon +\nonumber\\
&& + \left(-2 \log^2\,\varepsilon-\frac{2}{3} \pi ^2 \log\,\varepsilon-4 \log\,\varepsilon+\zeta_3-\frac{11 \pi ^4}{540}-\frac{5 \pi
   ^2}{6}\right) \varepsilon ^2+\cdots
\ea
The $\varepsilon^2$ term is correct in the logarithmically enhanced and transcendental terms, but a mismatch appears in the rational part. 
Extending the calculation at higher orders in the $\varepsilon$ expansion one gets the mismatch
\be
\label{eq:twist3mismatch}
\gamma_{3, \Delta-{\rm method}}-\gamma_3 = -\varepsilon ^2+\varepsilon ^3-\frac{5 \varepsilon ^4}{8}+\frac{\varepsilon ^5}{4}-\frac{23 \varepsilon ^6}{216}+O\left(\varepsilon ^7\right)
\ee
in quite close analogy with the twist-2 case.

\subsection{Improved expansion}

We illustrated the improved expansion of the Baxter equation in full details for the twist-2, 2 loop case. Here, we just show what happens at order ${\cal O}(\varepsilon^2)$
to show that a quite similar mechanism occurs. In particular, we want to give the explicit anomalous part in 
\be
F(z) = F_{\rm reg}(z) + F_{\rm anom}(z).
\ee
At three loops, the relevant terms contributing at ${\cal O}(\varepsilon^2)$ are 
\be
F_{\rm anom}(z) = \left(F_{\rm anom}^{(0)}(z)+\lambda\,F_{\rm anom}^{(1)}(z) +\lambda^2\, F_{\rm anom}^{(2)}(z)\right)\,\varepsilon^{4\,z} + \cdots .
\ee
Here are the explicit expressions. The one-loop term is simply
\be
F_{\rm anom}^{(0)}(z) = \frac{16^{-z}  \Gamma (-z) \Gamma \left(z+\frac{1}{2}\right)^3}{\Gamma \left(\frac{1}{2}-z\right)^3 \Gamma (z)}.
\ee
The two loop term is 
\ba
F_{\rm anom}^{(1)}(z) &=& \frac{2^{1-4 z} \pi ^5 \sec ^4(\pi  z)}{\Gamma \left(\frac{1}{2}-z\right)^6 \Gamma (z) \Gamma (z+1)}\,\log\varepsilon+\\
&& -\frac{3 2^{-4 z-1} \pi ^4 (2 z-1) \left(\psi ^{(1)}\left(\frac{3}{2}-z\right) \left(1-4 z^2\right)^2-\psi ^{(1)}\left(z+\frac{3}{2}\right) \left(1-4
   z^2\right)^2+32 z\right)}{(2 z+1)^5\,\sin(\pi  z)\, \cos^3(\pi  z) \Gamma \left(-z-\frac{1}{2}\right)^3 \Gamma \left(\frac{3}{2}-z\right)^3 \Gamma (z) \Gamma (z+1)
   } \nonumber
\ea
Finally, the three loop term is rather complicated, but has the following simple  expansion around $z=1/2$,
\be
F_{\rm anom}^{(2)}(z) = \log\,\varepsilon+\left(z-\frac{1}{2}\right)\, \left(\log^2\,\varepsilon-2\, \log\,\varepsilon\right) + {\cal O}\left((z-1/2)^2\right)
\ee
The three loop anomalous dimension is given by (recall $u=i\,z$)
\ba
\gamma_{3, \rm anom} &=& {F_{\rm anom}^{(0)}}' + \lambda\,\left({F_{\rm anom}^{(1)}}'-\frac{1}{4}\,{F_{\rm anom}^{(0)}}'''\right) + \\
&& + 
\lambda^2\,\left({F_{\rm anom}^{(2)}}'-\frac{1}{4}\,{F_{\rm anom}^{(1)}}'''+\frac{1}{48}\,{F_{\rm anom}^{(0)}}'''''\right)\nonumber
\ea
Evaluating the derivatives in $z=1/2$ we find at one-loop
\ba
{F_{\rm anom}^{(0)}}' &=& 0, \\
{F_{\rm anom}^{(0)}}''' &=&  3\,\varepsilon^2, \\
{F_{\rm anom}^{(0)}}''''' &=& 240\,\varepsilon^2\,(1-2\,\log\varepsilon+2\,\log^2\varepsilon).
\ea
At two-loops
\ba
{F_{\rm anom}^{(1)}}' &=& \frac{3}{4}\,\varepsilon^2, \\
{F_{\rm anom}^{(1)}}''' &=& 6\,\varepsilon^2\,(3-8\,\log\,\varepsilon+10\,\log^2\varepsilon).
\ea
Finally, at three loops
\be
{F_{\rm anom}^{(2)}}' = \varepsilon^2\,\log\varepsilon\,(-2+5\,\log\varepsilon).
\ee
Summing all the contributions we see that all logarithms cancel with a full result
\be
\gamma_{3, {\rm anom}} = +\varepsilon^2 + \cdots
\ee
which is precisely the term needed to correct the ${\cal O}(\varepsilon^2)$ mismatch in \refeq{eq:twist3mismatch}.

\section{Conclusions}

We have considered the minimal anomalous dimension of twist-2 and 3 operators in the 
$\mathfrak{sl}(2)$ sector of ${\cal N}=4$ SYM. For this quantity, three loop conjectures have been proposed for the 
closed function $\gamma(N)$, $N$ being the spin. These conjectures have never been proved. In all cases, the guessed expressions
can be expanded at large spin as
\be
\gamma(N) = \sum_{n=0}^\infty\sum_{m=0}^{M_n} a_{n,m}(\lambda)\,\frac{\log^m\,N}{N^n}.
\ee
These expansions are non trivial. For instance, the various terms obey generalized 
Moch-Vermaseren-Vogt (MVV) relations~\cite{Moch:2004pa,Vogt:2004mw} related to reciprocity 
properties~\cite{Basso:2006nk,Dokshitzer:2006nm,Beccaria:2007bb}.

\medskip
We have shown that the above expansion can be obtained in an algorithmic way from the long-range Baxter equation
valid in this sector. The method is based on an asymptotic expansion of the logarithm of the Baxter function $F(z) = \log\,Q(i\,z)$ suitable
in the large $N$ limit in a neighbourhood of $z = {\cal O}(N^0)$. Generically, it is composed of two parts
\be
F(z) = F_{\rm reg}(z) + F_{\rm anom}(z),
\ee
where $F_{\rm reg}(z)$ has a standard logarithmic expansion in $1/N$ and $F_{\rm anom}$ contains terms of the form 
\be
F_{\rm anom}(z) = \sum_{m=1}^\infty\sum_{n=0}^\infty \left(\frac{1}{N}\right)^{4\,m\,z+n}\,\sum_{k=0}^{K_{n,m}}\,a_{m,n,k}(z)\,\log^k\,N.
\ee

Our modest technical development completely bypasses the solution of the Bethe Ansatz equation and the 
determination of the large spin Bethe root density. Also, it does not rely on any conjecture about $\gamma(N)$. 
In principle, the method can be applied to more complicated examples where a Baxter formulation is available, possibly in nested form.
A nice example is the $\mathfrak{sl}(2|1)$ sector~\cite{Belitsky:2007zp} where a general Ansatz for $\gamma(N)$ is not known 
and it is an open question to prove the validity of reciprocity relations.

\acknowledgments
We thank G.~F.~De~Angelis for discussions and useful comments.

\end{document}